# Computational Investigation of Reactivity Parameters, UV-Vis and IR Spectra, NLO Properties, and Temperature-Dependent Thermodynamic Characteristics of Schiff-Based Interdigitated 5O.m (m=14,16) Liquid Crystalline Compounds: A DFT Analysis

Kritika Garg [1], Adrish Chakraborty[2], Ayon Bhattacharjee[2], Sandip Paul Choudhury[3], Sunita Kumari[4*], Debanjan Bhattacharjee[1*]

[1]*Department of Physics, Manipal University Jaipur, Jaipur, 303007*

[2]*Department of Physics, National Institute of Technology Meghalaya, Shillong, 793003*

[3]*Amity School of Applied Physics, Amity University Rajasthan, Jaipur, 303002*

[4]*Department of Physics, Indian Institute of Technology, Jodhpur, 342037*

**Abstract:** The article studies the different physical, vibrational, nonlinear optical and thermodynamical properties of higher homologs 5O.m (m = 14,16) liquid crystalline compounds using density functional theory. The optimized structure of 5O.m (m= 14,16) liquid crystals were obtained by using density functional theory (DFT) with B3LYP functional and standard basis set 6-31G (d, p). The Infrared spectra (IR), various physical properties such as HOMO-LUMO, nonlinear optical properties (NLO), reactivity parameters, relative energy gaps, and electrostatic potential function are computed and analysed using the optimised structure of 5O.m liquid crystal. The time dependent density functional theory (TD-DFT) has been used to analyse and obtain UV-Vis spectra for both LC compounds. It is observed that the 5O.m (m=14,16) liquid crystals are showing lower value of HOMO-LUMO energy gaps as 4.17ev which resulted in some highly fascinating optical and physical properties. Using DFT excellent agreement is observed between all spectrum patterns and the simulated UV-Vis and IR spectra. This article however, also discussed about temperature dependent thermodynamical properties such as zero-point vibrational energy (ZPVE), total thermal energy, total specific heat capacity, rotational constants and total entropy which enable us to understand the phase transition behaviour and specific transition temperatures of different phases.

---

[*]Corresponding author.

Email address: bhattacharjee8@gmail.com (D. Bhattacharjee), sunitabhu@iitj.ac.in (S Kumari)

# 1.Introduction

Liquid crystals have played a significant part in application devices throughout the last few decades. Liquid crystals have the potential for such applications due to their unique mesogenic behaviour. Because of its mesogenic character, it is an exceptional material in the domains of optics, electro optics, thermal devices, biosensors, and switching devices and among others [1]. To investigate the more distinctive aspects of the LC phase, it is required to better understand the structural and dynamical properties, as well as the correlation of the mesogenic phases with temperature and chemicals. In general, the shape and size of the molecule influence the property and phase transition behaviour of LC substances [2]. Intra and inter molecular interactions alter the molecular property of the molecule during the LC phase. As a result, vibrational spectroscopy has a high potential for understanding the compound's molecular dynamics. To conduct a more thorough investigation of the compounds infrared spectroscopy also important in the LC field [3][4][5]. Based on molecular behaviour, IR helps to understand the dynamics of the molecules. and DFT computational tools aid in this understanding [6].

LC materials are of great interest due to their absorption characteristics in certain wavelength bands. Absorption wavelengths and oscillator strengths have been calculated using a variety of approaches, including density functional theory (DFT). However, DFT's ability to handle organic compounds is limited. As a result, semi-empirical approaches are widely employed for simulation of UV-Vis spectrum of LC materials. These approaches allow for the computation of electronic transitions and corresponding radiation energy. In this study TD-DFT technique has been used to compute UV-Vis spectra of 5O.m(m=14,16) liquid crystal which gives insights to the absorption bands, oscillator strength etc of liquid crystalline compounds [7][8][9][10].

The computational techniques used to investigate LC molecules have seen tremendous modification in recent decades. Several computational techniques with various basis sets have been tested this year. Therefore, it appears from several sources of literature that density functional theory (DFT) combined with the B3LYP approach and the conventional basis set of 6-31G(d,p) produces findings that are adequate and appropriate for various liquid crystalline compounds [11]. This article examines the molecule conformation and isoelectronic density surface with electrostatic potential distribution using a density functional theoretical (DFT) technique. Additionally, the relative energy gap, atomic orbital composition, and absolute energy of the compounds were assessed using the DFT method. Information from HOMO and LUMO description is used to create new molecules and enhance the properties of liquid crystalline compounds [3][12][13][14][15][16].

Thermodynamical properties with the variation of temperature of liquid crystal are influenced by their anisotropic nature which means they have different physical properties in different directions. Due to the phase transition in liquid crystal their specific heat energy changes which indicates the absorption or release of energy. Thermal energy, specific heat and entropy of liquid crystal are closely tied to their unique mesophase transitions [17][6][18][19][20][21].

DFT simulations are used in this study to examine the NLO characteristics and reactivity parameters of liquid crystals. DFT's anticipatory capability to grasp the underlying principles governing NLO behaviour in various liquid crystal devices. A detailed examination of these properties will improve the understanding of liquid crystals. These understanding also gives designers of modern photonic devices with new ideas for improving their usefulness and efficiency. Polarizability and hyperpolarizability in liquid crystals are influenced by elements such as molecular shape, size, and arrangement within the material. By controlling these factors, researchers may create liquid crystal materials with exact optical properties, making them useful in several applications such as liquid crystal displays (LCDs) and electro-optic devices [3][19][22][23] [24].

Understanding the polarizability and hyperpolarizability of liquid crystals are essential for developing new technologies that take advantage of their distinctive properties. Researchers regularly employ methods like optical spectroscopic, electric field-induced birefringence measurements, and nonlinear optical measurements to characterize how liquid crystal materials respond to external electric fields and light. In this article, 5O.m (m=14,16) were used. These compounds are specially interdigitated and slightly bent like liquid crystal materials. The main focus in this work is to simulate and interpret the theoretical results of both compounds [11][25][26][27][28][14][29][30][31].

## 2.Experimental and computational details

The samples used in this article were interdigitated 5O.m (m=14, 16) liquid crystals and were synthesised reported elsewhere [32][18][33][34][35]. The phase transition temperatures of these two compounds are shown in Table.1. The optimized structure of the two liquid crystalline compounds were obtained using density functional theory (DFT) approach and it is shown in fig.1.

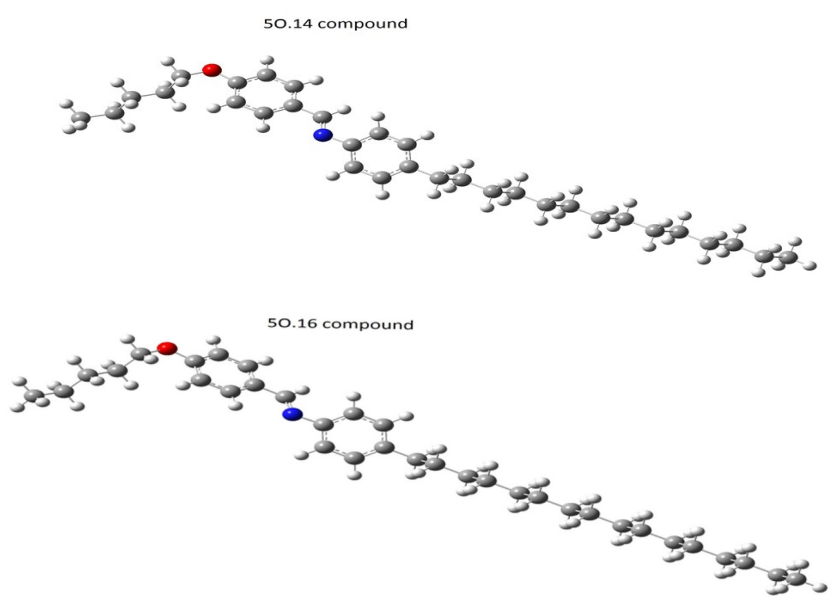

Fig.1. The optimized molecular structure of compounds 5O.14 and 5O.16.

Table 1 Phase transition temperature of 5O.14 and 5O.16 liquid crystal compounds.

| 5O.16 | Temperature | 5O.14 | Temperature |
|---|---|---|---|
| I-N | 69°C | I-N | 71.5°C |
| N-$S_A$ | 67.7°C | N-$S_A$ | 69.3°C |
| $S_A$-C | 54.6°C | $S_A$-$S_B$ | 51.7°C |
|  |  | $S_B$-C | 43°C |

## 2.1 Computational method

The Gaussian 09 software package was used to perform all the quantum chemical density functional calculations on two interdigitated liquid crystalline substances. "Valence double zeta plus single polarization and diffuse" basis set, 6-31G (d, p), is more accurate for barrier heights, energy changes in isogyric reactions, and conformational energies. Therefore the hybrid density functionals theory was used for this computation, with the conventional basis set 6-31G (d, p) paired with the Lee, Yang, and Parr (B3LYP) hybrid functional for the correlation component and the Becke three hybrid density functionals for the exchange part [36][37][38]. For all DFT calculations, the Gaussian 09 package of software is utilized. In this method, the second derivative of the polarizability with respect to the normal coordinates was taken. The IR activities were described using DFT. In this article Important nonlinear optical phenomena like dipole moment, polarizability and hyperpolarizability, reactivity parameters and thermodynamical properties with the variation of temperature are studied and analyzed by using DFT [12] [13] [15] [18] [19] [39] [40][9].

## 3. Results and discussions

3.1 HOMO-LUMO and Electrostatic potential (ESP) Study

The quantum mechanical density functional study provides further understanding of the LC compound to help with the precise understanding of molecular conformation and dynamics. The DFT is an effective method for studying the mentioned compounds like dipole moment, HOMO LUMO, relative energy gaps, and electrostatic potential distribution more precisely. The compounds that were employed were interdigitated and had unusual sorts of partially bent bends like those found in nature. It is evident from Fig.1. that these compounds are uniaxial. The entire structure is shown in fig.1. was created and optimized using density functional theory (DFT) and the Gaussian 09 program.

For successful application we have also used the B3LYP functional with standard basis set 6-31G (d, p) in both the compounds [36][41][15][42][3][43][44].

When these compounds were investigated for their electrostatic potential, the stated optimal molecular geometries of both were sterically bulky, as Fig.2. and 3 show. Furthermore, it was discovered that the reddish colour of the LC compounds' cores suggested that they possessed a negative electrostatic potential. The compound's increased electronegative potential in the C=N bond is indicated by the dark red colour in fig.2 and 3. The positive electrostatic potential of both compounds was represented by the sky-blue colour. The short terminal chains of both compounds included the majority of the positively charged atoms.

The compounds 5O.14 and 5O.16 have electrostatic potential energies of $-4.250e^{-2}$ to $4.250e^{-2}$ and $-4.29e^{-2}$ to $4.29e^{-2}$, respectively. This indicates that the chemical 5O.16 has a much greater negative energy due to the extra $CH_2$ molecule in the terminal chain. The total electrostatic potential study (ESP) revealed that the electron density was spread fairly equally in both compounds, which broadens the range of applications for LC compounds.

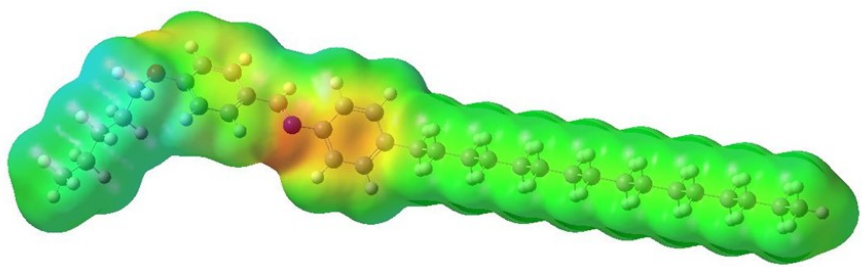

Fig.2. Electrostatic potential (ESP) structure of 5O.14 compound.

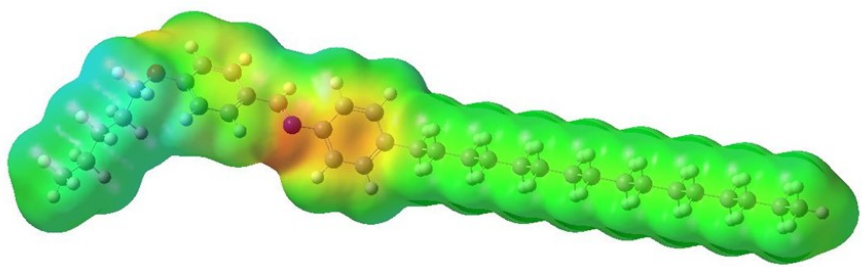

Fig.3. Electrostatic potential (ESP) structure of 5O.16 compound.

It was easy to investigate the atomic orbital composition, relative energy gap, and absolute energy of the compounds from the HOMO and LUMO studies. These findings offer important details regarding the optical and physical characteristics and contribute to improving the composition of these compounds. Fig.4, 5, and Table 2 respectively depict the two compound's three-dimensional HOMO and LUMO structures as well as their relative energies. HOMO-LUMO energy separation was used to calculate the compound's kinetic stability and molecular reactivity.

Between the highest occupied molecular orbital and the next lowest unoccupied molecular orbital, there is an energy gap known as the HOMO-LUMO energy gap. It has been observed that the energy gaps in liquid crystals are generally small, indicating interesting optical and electrical properties. Remarkable physical features result from molecules with small HOMO-

LUMO energy gaps, which are often stable and show lower-energy simulations. From Table 2 the energy gap of both the compounds is modest in comparison to usual situation. These compounds exhibit a variety of remarkable electronic properties, including thermo-excited intramolecular electron transport and metallic conductivity.

Alkyl chains in liquid crystal play a dominant role in molecular properties. An increment in the number of alkyl chain causes an increase in HOMO- LUMO energies and decrease in the energy gaps. An increase in the length of the alkyl chain is clearly preferred, as evidenced by the energy gap values of isolated single molecules [45][27][15][22][3][42][46][42][13][47].As table 2 both the compounds exhibit identical behaviour as discussed.

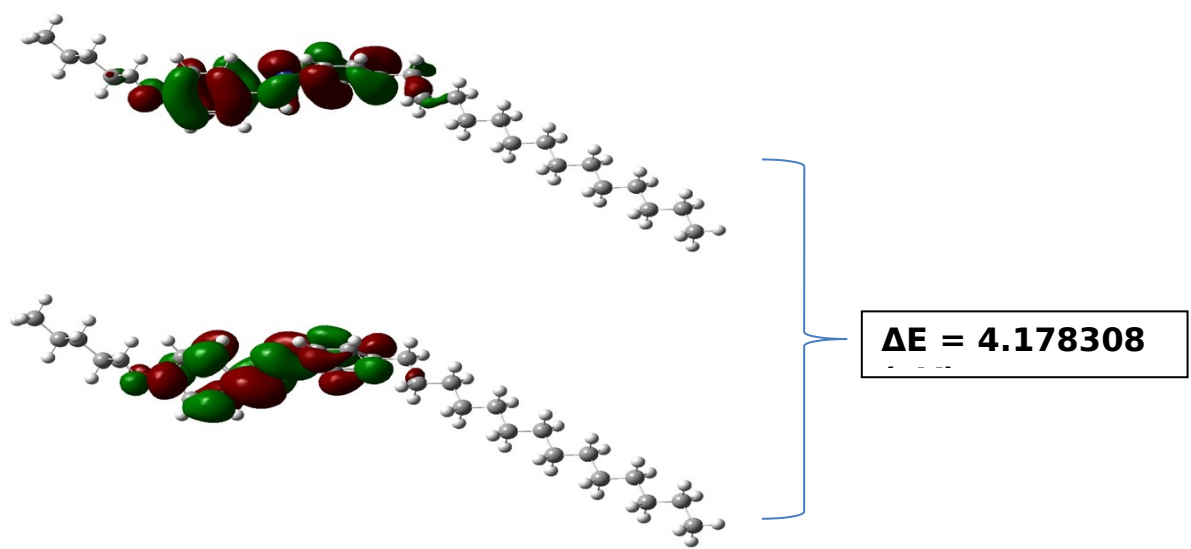

Fig.4.HOMO and LUMO of 5O.14 compound.

**ΔE = 4.178308**

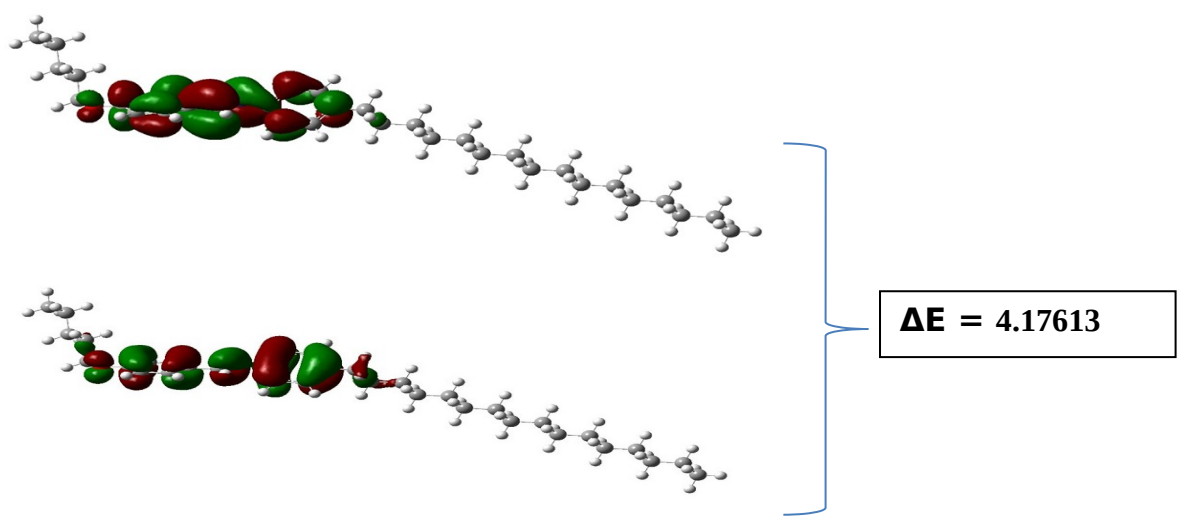

Fig.5.HOMO and LUMO of 5O.16 compound.

**ΔE = 4.17613**

Table 2 HOMO- LUMO and relative energy gap (ΔE) of both the compounds.

| Sample | HOMO | LUMO | ΔE(Hartree) |
|--------|----------|----------|-------------|
| 5O.14  | -0.19967 | -0.04612 | 0.15355     |
| 5O.16  | -0.19968 | -0.04621 | 0.15347     |

3.2 Thermodynamical parameters

Thermodynamical properties of liquid crystals such as Zero-point vibrational energy, Total thermal energy, total specific heat, and total entropy plays a prominent role for evaluating the types of liquid crystal and its different phases. Ordering of the liquid crystal varies with the molecular orientational temperature.  When a material starts melting, its phase changes from crystal to liquid crystal, and this process demands energy (endothermic) from its surroundings. Similarly, the crystallisation of a liquid crystal is an exothermic process that releases energy into the surrounding environment. In terms of structural change, the melting transition from a crystal to a liquid crystal is a highly abrupt phase transition with comparatively significant transition energy. A mesogenic material can undergo a number of transitions with corresponding entropy changes when heated from its crystalline state. The degree of internal order in the system correlates with the transition entropies between the various liquid crystalline states and the isotropic state [48][49][21][6][18][20][50][51][52].

The thermodynamical properties as Zero-point vibrational energy, Total thermal energy, total specific heat, total entropy, and rotational constants of 5O.m (m=14,16) liquid crystalline compounds are calculated and studied using DFT method.[53] These calculations were performed at different temperatures and 1atm pressure. For 5O.14 liquid crystalline compound these parameters are calculated at 323K, 342K and 343K temperatures and for 5O.16 liquid crystalline compound these parameters are calculated at 309K, 342K and 343K temperatures. The resulting data is shown in table 3 and table 4. It is observed form the comparative analysis of both the compounds that with the increasing temperature, the first parameter of table 3 and table 4 is zero-point vibrational energy (ZPVE) is constant which explains the minimum energy of molecular vibrations at absolute 0K temperature. The second parameter of table 3 and 4  is Total thermal energy which contains translational energy, electronic energy, rotational energy and vibrational energy whereas the vibrational energy makes the primary contribution to the total internal energy so with the increasing temperature vibrational energy increases as the molecules starts partially disordered in order to change phases from nematic to smectic A ,smectic A to smectic B and also smectic A to crystal as a result the total thermal energy increases with the increasing temperature. The third parameter is total specific heat capacity which refers to the released or absorbed amount of thermal energy to change the temperature by 1K of per 1 mol of 5O.14 and 5O.16 both compounds so it is observed from table 3 and 4 the total specific heat increases with the increasing temperature.[10] The fourth parameter of table 3 and table 4 is total entropy which refers to the randomness and disorderness of both liquid crystalline compounds as from the table 3 and table 4 total entropy increases with the increasing temperature which shows that with the variation of temperature molecules undergo phase changes in order to increase randomness within the compounds. The fifth parameter of table 3 and table 4 is rotational constants A, B, C which is constant with the increasing temperature this gives understanding that the 5O.14 and 5O.16 compounds have précised molecular symmetry and structure. It was discovered that the compound 5O.16 had greater thermal energy values than the compound 5O.14, which was due to the compound exhibiting more electro-negativity than the 5O.14 compound [44][43][3][54][55][56].

Table 3 Thermodynamic parameters of 5O.14 liquid crystalline compound.

| Sample 5O.14 | | | | |
|---|---|---|---|---|
| Parameters | | | Temperature | |
| | | 323 K | 342K | 343K |
| Zero-point vibrational energy (ZPVE) (in *kcal/mol*) | | 469.41706 | 469.41706 | 469.41706 |
| Thermal Energy (E) (in *kcal/mol*) | Elec. | 0.000 | 0.000 | 0.000 |
| | Vib. | 495.200 | 498.216 | 498.061 |
| | Rot. | 0.963 | 1.022 | 1.019 |
| | Trans. | 0.963 | 1.022 | 1.019 |
| | Total | 497.126 | 500.261 | 500.100 |
| Total specific heat capacity ($C_v$) (in *kcal/mol/K*) | | 152.255 | 161.284 | 160.833 |
| Total entropy (S) (in *kcal/mol/K*) | | 262.221 | 271.756 | 271.280 |
| Rotational constants (in GHz) | A | 0.31497 | 0.31497 | 0.31497 |
| | B | 0.01263 | 0.01263 | 0.01263 |
| | C | 0.01254 | 0.01254 | 0.01254 |

Table 4. Thermodynamic parameters of 5O.16 liquid crystalline compound

| Sample 5O.16 | | | |
|---|---|---|---|
| Parameters | | Temperature | |
| | 309 K | 342 K | 343K |
| Zero-point vibrational energy (ZPVE) (in *kcal/mol*) | 505.31675 | 505.31675 | 505.31675 |

| | | | | |
|---|---|---|---|---|
| Thermal Energy (E) (in *kcal/mol*) | Elec. | 0.000 | 0.000 | 0.000 |
| | Vib. | 529.282 | 536.092 | 536.258 |
| | Rot. | 0.889 | 1.019 | 1.022 |
| | Trans. | 0.889 | 1.019 | 1.022 |
| | Total | 531.059 | 538.131 | 538.303 |
| Total specific heat capacity ($C_v$) (in *kcal/mol/K*) | | 150.741 | 171.827 | 172.309 |
| Total entropy (S) (in *kcal/mol/K*) | | 264.595 | 286.962 | 287.470 |
| Rotational constants (in GHz) | A | 0.26689 | 0.26689 | 0.26689 |
| | B | 0.01014 | 0.01014 | 0.01014 |
| | C | 0.00997 | 0.00997 | 0.00997 |

3.3 NLO and Reactivity analysis

3.3.1 Nonlinear optical properties (NLO)

As published work[44][43][3][54][55], both the compounds 5O.14 and 5O.16 were employed in this study have some notable optical properties, which are listed in Table 5 that displays the dipole moment of the compounds calculated from the three Cartesian directions. The dipole moment of the long axis, on the other hand, is substantially bigger than that of the other two axes. In electro-optic systems and displays, the director reorientation in electric fields is determined by the dipole moment of liquid crystals. Understanding the molecule's energy gap is essential for determining its stability and reactivity. Higher levels of a compound's energy gap ($E_g$) frequently exhibit solid-like behaviour. The energy gap of liquid crystal compounds in the nO.m series is typically smaller. Lower gap values interact with molecules with less energy, making them simpler to polarize with electric and magnetic fields and this has lots of uses that are quite advantageous. As table 5 shows that both the compounds have relatively good dipole moment which shows easily polarizability and molecular conformation for both 5O.14 and 5O.16 compounds.

Nonlinear optics (NLO)-based innovative applications include optical computing, optical storage, and the generation of higher-order harmonic signals, among others. In the presence of a sufficiently intense light source, molecules with anisotropic polarizability undergo a transition from a state of high potential energy to a state of low potential energy. Major structural criteria for a molecule to exhibit optical nonlinearity is the presence of a bridge separating an electron donor group and an electron acceptor group.

Liquid crystalline molecules display strong optical nonlinear effects due to their elongated conjugated molecular structure and proclivity to align in the direction of an applied electric field. The B3LYP/6-31G (d, p) level components of the polarizability and hyperpolarizability tensors are computed here. Table 5 shows the total molecule dipole moment, total polarizability, asymmetry parameters, anisotropic polarizability, and first-order hyperpolarizability calculated using DFT [3][44][43][19][54][22][57][23][24].

Table 5 displays the components of the polarizability and hyperpolarizability tensors in addition to the asymmetry parameter (η), isotropic polarizability ($\alpha_{iso}$), anisotropic polarizability (Δα), total molecular dipole moment ($\mu_{total}$), total polarizability ($\alpha_{total}$), and first order hyperpolarizability ($\beta_o$). B3LYP functionals were used to carry out the computation. The statistics in Table 5 demonstrate that the components of dipole moment, polarizability, and hyperpolarizability have greater magnitudes. The compound's smaller energy gap explains this. The molecule then exhibits improved NLO characteristics, which are determined by the following formulas [28][58][59][22][60][42] [27].

$$\mu_{total} = \sqrt{\mu_x^2 + \mu_y^2 + \mu_z^2} \tag{1}$$

$$\alpha_{total} = \frac{1}{\sqrt{2}} \times \sqrt{(\alpha_{xx} - \alpha_{yy})^2 + (\alpha_{yy} - \alpha_{zz})^2 + (\alpha_{zz} - \alpha_{xx})^2 + 6\alpha_{xy}^2 + 6\alpha_{xz}^2 + 6\alpha_{yz}^2} \tag{2}$$

$$\eta = \frac{\alpha_{xx} - \alpha_{zz}}{\alpha_{xx} - \alpha^{iso}} \tag{3}$$

$$\Delta\alpha = \alpha_{xx} - \frac{\alpha_{yy} + \alpha_{zz}}{2} \tag{4}$$

$$\alpha^{iso} = \frac{\alpha_{xx} + \alpha_{yy} + \alpha_{zz}}{3} \tag{5}$$

$$\beta_x = \beta_{xxx} + \beta_{xyy} + \beta_{xzz} \tag{6}$$

$$\beta_y = \beta_{yyy} + \beta_{xxy} + \beta_{yzz} \tag{7}$$

$$\beta_z = \beta_{zzz} + \beta_{xxz} + \beta_{yyz} \tag{8}$$

$$\beta_0 = \sqrt{\beta_x^2 + \beta_y^2 + \beta_z^2} \tag{9}$$

Table 5 NLO properties of both 5O.14 and 5O.16 compounds.

| | Dipole moment (μ) (in Debye) | | | |
|---|---|---|---|---|
| Compound | $\mu_x$ | $\mu_y$ | $\mu_z$ | $\mu_{total}$ |
| 5O.14 | -1.7719 | -0.6361 | 0.0335 | 1.8829 |
| 5O.16 | -1.7747 | 0.6184 | 0.1259 | 1.8835 |

| Polarizability (in Debye-Å) | 5O.14 | 5O.16 |
|---|---|---|
| $\alpha_{xx}$ | 566.2072 | 642.8741 |
| $\alpha_{yy}$ | 347.8296 | 318.5029 |
| $\alpha_{zz}$ | 237.2241 | 295.2050 |
| $\alpha_{xy}$ | 3.146086 | -49.6685 |
| $\alpha_{xz}$ | -41.62181 | 51.5643 |
| $\alpha_{yz}$ | 21.36635 | 23.0135 |
| $\alpha^{iso}$ | 383.7536 | 418.8606 |
| $\Delta\alpha$ | 273.6803 | 336.0201 |
| $\eta$ | 1.8031 | 1.5520 |
| $\alpha_{total}$ | 425.8462 | 510.4566 |
| Hyperpolarizability (in Debye-Å$^2$) | 5O.14 | 5O.16 |
| $\beta_{xxx}$ | -537.3413 | 679.1230 |
| $\beta_{yyy}$ | 28.1715 | -17.7174 |
| $\beta_{zzz}$ | -37.2241 | 40.7202 |
| $\beta_{xyy}$ | -44.8925 | 127.5892 |
| $\beta_{xxy}$ | 90.4198 | -624.4152 |
| $\beta_{xxz}$ | -190.8304 | 150.9691 |
| $\beta_{xzz}$ | 71.1142 | 29.6806 |
| $\beta_{yzz}$ | -16.5777 | -18.8536 |
| $\beta_{yyz}$ | -44.5158 | 13.7621 |
| $\beta_{xyz}$ | 40.9223 | -138.5751 |
| $\beta_x$ | -511.1196 | 836.3928 |
| $\beta_y$ | 102.0136 | 660.9862 |
| $\beta_z$ | 272.5703 | 205.4514 |
| $\beta_0$ | 588.1705 | 1085.6638 |

3.3.2 Reactivity parameters analysis

It is observed that both 5O.14 and 5O.16 compounds show lower energy gap values which shows easily interaction with other molecules with minimum energy. Some chemical reactivity parameters such as Ionization Potential (I), Electron Affinity (A), Electron Negativity (χ), Global Hardness (η), Global Softness (S), Chemical Potential (μ), Global electron philicity Index(ω) are shown in Table 6.The findings reveal that the reactivity parameters of the 5O.16 liquid crystalline compound are marginally greater than those of the 5O.14 liquid crystalline compound, due to an increase in the alkyl chain thus the chemical reactivity parameters give understanding to the electronic structure, chemical behaviour of both the compounds [61][62].The reactivity parameters which is shown in Table 6, are calculated with the given relations:

$$I = -E_{HOMO} \quad (10)$$

$$A = -E_{LUMO} \quad (11)$$

$$\mu = \frac{-(I+A)}{2} \quad (12)$$

$$\eta = \frac{I-A}{2} \quad (13)$$

$$\omega = \frac{\mu^2}{2\eta} \quad (14)$$

$$S = \frac{1}{\eta} \quad (15)$$

$$\chi = -\mu \quad (16)$$

Table 6 Reactivity Parameters for both 5O.14 and 5O.16 compounds.

| Reactivity Parameters | 5O.14 | 5O.16 |
|---|---|---|
| Ionization Potential (I) (in Hartree) | 0.19967 | 0.19968 |
| Electron Affinity (A) (in Hartree) | 0.04612 | 0.04621 |
| Electron Negativity (χ) (in Hartree) | 0.122895 | 0.122945 |
| Global Hardness (η) (in Hartree) | 0.076775 | 0.076735 |
| Global Softness (S) (in Hartree$^{-1}$) | 13.02507 | 13.03186 |
| Chemical Potential (μ) (in Hartree) | -0.122895 | -0.122945 |
| Global electron philicity Index(ω)(in Hartree) | 0.098360 | 0.0984913 |

3.4 UV-Vis spectra analysis

Electronic transitions involving π and/or n electron systems are closely related to molecular UV spectra. Particularly, compounds that include aromatic groups have strong UV absorption properties. Clarifying the electrical structure of liquid crystal (LC) molecules and pushing the boundaries of optical device technology need a thorough understanding of the interactions between light and LC molecules. Quantum chemical approach is essential to further explore the basic charge and excitation energy transfer mechanisms in low and high absorption liquid crystals. These fundamental procedures are crucial in establishing how effective these systems are for a range of uses, such as solar cells and flexible displays [63][10][8]. Time dependent density functional theory (TD-DFT) at B3LYP functional with standard basis set 6-31G(d,p) has been used to analyse the UV-Vis spectra of both the 5O.14 and 5O.16 compounds. The UV-Vis spectra of both 5O.14 and 5O.16 liquid crystalline compounds shown in figure 6(a) and 6(b) respectively. It is discovered that for 5O.14 liquid crystalline compound three absorption peaks occurred in the UV region at 343.05nm($\lambda_1$), 293.38nm ($\lambda_2$), 282.55nm ($\lambda_3$) and oscillator strength corresponding to $\lambda_1$, $\lambda_2$, and $\lambda_3$ is 0.50, 0.31, 0.074 respectively. For 5O.16 liquid crystalline compound it is observed that three absorption peaks occurred at 335.67nm($\lambda_1$), 286.97nm ($\lambda_2$), 271.22nm ($\lambda_3$) and oscillator strength corresponding to $\lambda_1$, $\lambda_2$, and $\lambda_3$ is 0.69, 0.48, 0.002 respectively. These results shows that no absorption has been observed in the visible region. These peaks arise due to HOMO LUMO transitions. The strongest band appears with absorption maxima ($\lambda_{max}$) for 5O.14 LC compound at 343.05nm and for 5O.16 LC compound at 335.67nm. Higher oscillator strength as a result of these transitions is associated with the strongest absorption bands at $\lambda_1$ [8][64][43][65].

Table 6 Theoretical UV-Vis spectra values (wavelength, excitation energy, oscillator frequency) and MO contributions of both 5O.14 and 5O.16 LC compounds.

| Molecule | Wavelength (nm) | Excitation Energy(eV) | Oscillator strength(*f*) | MO contribution |
|---|---|---|---|---|
| 5O.14 | 343.05 | 3.6141 | 0.5005 | HOMO→LUMO (83.48%) |
|  | 293.38 | 4.2260 | 0.3190 |  |
|  | 282.55 | 4.3880 | 0.0741 | HOMO-2→LUMO (33.08%) |
|  |  |  |  | HOMO-1→LUMO (64.72%) |
| 5O.16 | 335.67 | 3.6936 | 0.6908 | HOMO→LUMO (84.03%) |
|  | 286.97 | 4.3205 | 0.4892 |  |
|  | 271.22 | 4.5713 | 0.0029 | HOMO-1→LUMO (69.35%) |

|  | HOMO-2→LUMO (75.06%) |
| --- | --- |

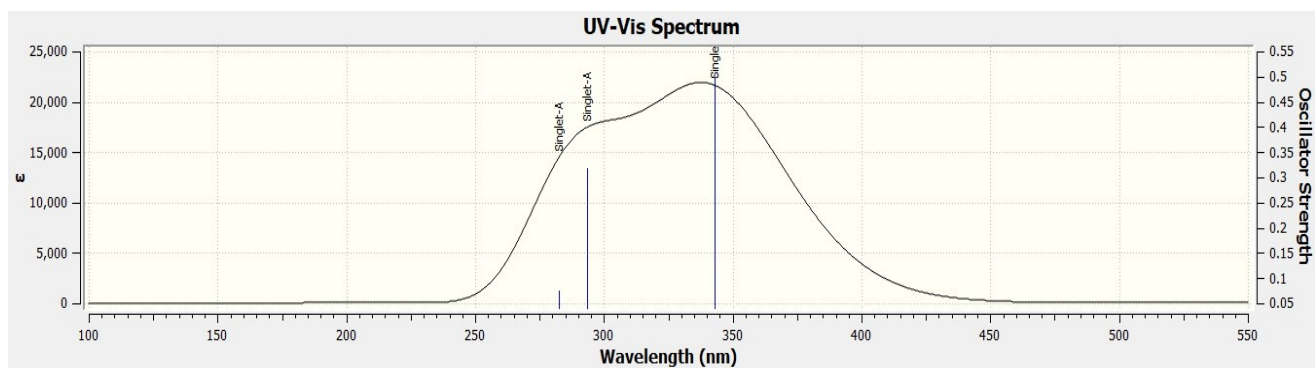

Fig. 6(a). DFT generated UV-Vis spectra of 5O.14 liquid crystalline compound.

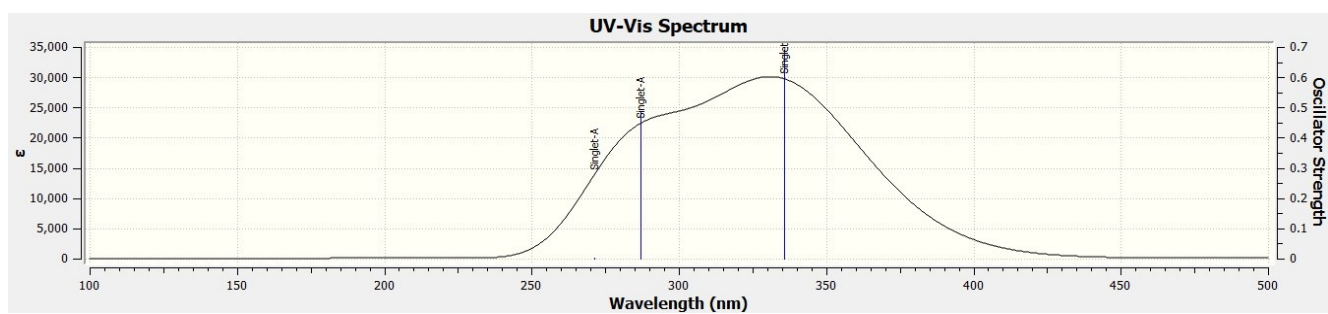

Fig.6(b). DFT generated UV-Vis spectra of 5O.16 liquid crystalline compound.

3.5 IR spectra study

Frequency assignment discussion

A careful explanation of the fig.6. reveals an anomalous behaviour of the spectral features for all IR bands. The IR spectra of both 5O.14 and 5O.16 LC compounds are shown in fig. 6.

In most of the liquid crystal, the presence of the alkyl chain and core of the aromatic benzene ring plays an important role for the spectroscopic behaviour of the compounds [44][3][66][59][67][19][5][22][68][49][69][22][36][38][9].

Table 7 Frequency assignment for 5O.14 and 5O.16 compounds.

| | Frequency Assignments | | | |
| --- | --- | --- | --- | --- |
| Serial No. | Band Position (referred) | Assignment. | 5O.14(IR) | 5O.16(IR) |
| 1. | 3000-2850 | C-H Stretching mode. (Strong) | 3106.69  3077.31  3030.91 | 3110.56  3077.31  3027.82 |

| 2. | 2000-1650 | Substituted benzene rings. (weak) | 1699.31 | 1696.22 |
| --- | --- | --- | --- | --- |
| 3. | 1690-1640 | C=N stretching mode. (strong) | 1669.93 | 1669.92 |
| 4. | 1600-1400 | Quadrant stretching mode of the aromatic ring. (Medium, weak, multiple bands) | 1652.92 1557.03 | 1649.82 1557.63 |
| 5. | 1480-1350 | C-H in plane bending mode. (variable) | 1361.39 | 1364.48 |
| 6. | 1300-1000 | C-O anti symmetric stretching mode. (Two bands or multiple bands | 1318.08 1291.79 1268.59 | 1314.99 1291.79 1238.59 |
| 7. | 1200-1000 | Aromatic C-H in plane bending mode. | 1188.94 1066.77 | 1188.94 1063.67 |

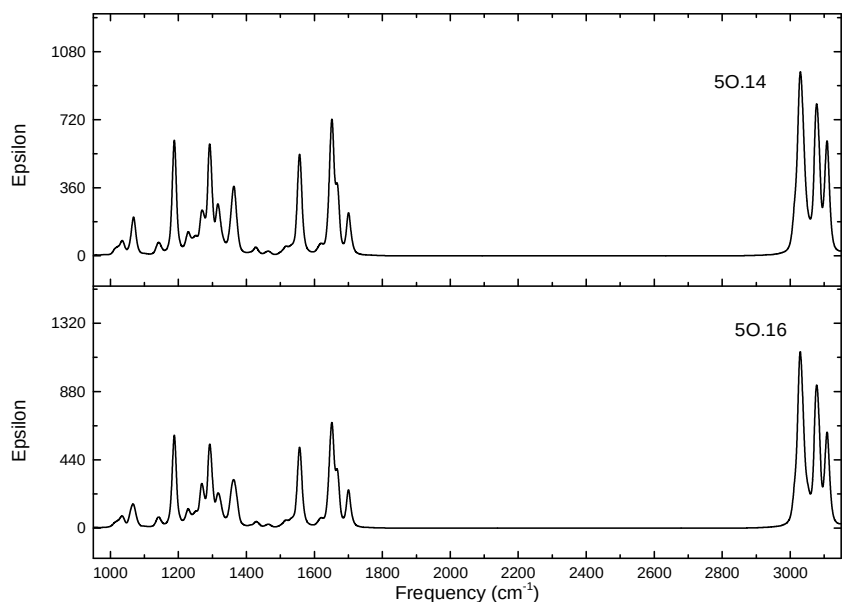

Fig.7. IR spectra of both 5O.14 and 5O.16 compounds.

The observed bends in the IR spectrum, which occurred in the range of 3000 cm$^{-1}$ to 3100 cm$^{-1}$, were caused by the alkyl chains in both compounds. These vibrations had a stretching mode and a medium intensity. The IR graph at the regions 1699.31 cm$^{-1}$, 1669.93 cm$^{-1}$ for 5O.14 compound and 1696.22 cm$^{-1}$, 1669.92 cm$^{-1}$ for 5O.16 compound were because of the benzene ring and C=N respectively in both the compounds. The stretching mode of these vibrations was strong, as were the intensities. Due to the aromatic's quadrant stretching modes, two different peaks were found at the bend position of 1653 cm$^{-1}$ to 1556 cm$^{-1}$ in both compounds. The C-H bending mode of the compounds 5O.14 and 5O.16 caused the bend at 1361.39 cm$^{-1}$ and 1364.48 cm$^{-1}$, respectively, and the mode of intensities was varying at this point. The ester effect (C-O-C) was also observed at the position 1320 cm$^{-1}$ to 1265 cm$^{-1}$ range in both compounds. In that particular region, both the stretching mode of vibration and the intensities' impacts were strong or extremely powerful. The peaks positions at 1189 cm$^{-1}$ to and 1060 cm$^{-1}$ in both the compounds were indicating the aromatic C-H in plane bending modes.

Above IR spectroscopic analysis shows that both compounds are IR active. Because of it, a clear and understanding of the molecular dynamics and spectral behaviour of the compounds with temperature was observed and understood [44][3][66][59][67][19][5][22][68][49][69][22].

**4.Conclusions**

The molecules 5O.14 and 5O.16 were subjected to quantum computational analysis utilizing the density functional theoretical method. The computations were carried out using the standard basis set 6-31G (d, p) in conjunction with the Becke three hybrid density functionals for the exchange component and the Lee, Yang, and Parr hybrid functional for the correlation part (B3LYP). The electrostatic potential function revealed that the negatively charged aromatic core of these compounds was dispersed very uniformly.

The vibrational spectra, nonlinear optical properties, dipole moment, thermodynamical properties and HOMO-LUMO energies of 5O.m(m=14,16) compounds were examined using density functional theory. The HOMO-LUMO energy gap ΔE=0.15 was seen to be fairly narrow. This little energy gap makes it very simple to improve the molecular characteristics for better application. Using this functional basis set, the simulated spectra for the UV-Vis and IR result demonstrates great agreement and both compounds were IR active, which aids in the accurate investigation of molecular dynamics. Additionally, it has been noted that both the compounds show strong nonlinear optical effect due to its high static hyperpolarizability. The NLO analysis, dipole moment analysis and electrochemical thermal analysis contribute to developing a new unique optical based liquid crystalline material. This study helps in generating new optoelectrical devices and advanced liquid crystalline materials.

## 5.References